\documentclass[twocolumn,superscriptaddress,prb,10pt]{revtex4-1}
\usepackage{amsmath,amssymb}
\usepackage{tikz}
\usepackage{graphicx}
\usepackage{color}
\usepackage[colorlinks,bookmarks=false,citecolor=blue,linkcolor=red,urlcolor=blue]{hyperref}
\usepackage{times}

\newcommand*\circled[1]{\tikz[baseline=(char.base)]{
            \node[shape=circle,draw,inner sep=2pt] (char) {#1};}}

\begin{document}

\title{Simulating the Generalized Gibbs Ensemble (GGE): a Hilbert space 
Monte Carlo approach} 

\author{Vincenzo Alba}
\affiliation{International School for Advanced Studies (SISSA),
Via Bonomea 265, 34136, Trieste, Italy, 
INFN, Sezione di Trieste}

\date{\today}

\begin{abstract} 

By combining {\it classical} Monte Carlo and Bethe ansatz techniques we devise a numerical 
method to construct the Truncated Generalized Gibbs Ensemble (TGGE) for the spin-$\frac{1}{2}$ 
isotropic Heisenberg ($XXX$) chain. The key idea is to sample the Hilbert space of the model 
with the appropriate GGE probability measure. The method can be extended to other integrable 
systems, such as the Lieb-Liniger model. We benchmark the approach focusing on 
GGE expectation values of several local observables. 
As finite-size effects decay exponentially with system size, moderately large chains are sufficient to 
extract thermodynamic quantities. The Monte Carlo results are in  
agreement with both the Thermodynamic Bethe Ansatz (TBA) and the Quantum Transfer Matrix approach (QTM). 
Remarkably, it is possible to extract in a simple way the steady-state 
Bethe-Gaudin-Takahashi (BGT) roots distributions, which encode complete information about the GGE 
expectation values in the thermodynamic limit. Finally, it is straightforward to simulate extensions 
of the GGE, in which, besides the local integral of motion (local charges), one includes {\it arbitrary} 
functions of the BGT roots. As an example, we include in the GGE the first non-trivial quasi-local 
integral of motion.

\end{abstract}

% \pacs{73.43.Cd, 71.10.Pm  {\tt check!}}

\maketitle

%#############-- INTRODUCTION --########################################
\paragraph*{Introduction.---}

The issue of how statistical ensembles arise from the out-of-equilibrium 
dynamics in {\it isolated} quantum many-body system is still a fundamental, 
yet challenging, problem. The main motivation for the renewed interest 
in this topic is the high degree of control reached in out-of-equilibrium 
experiments with cold atomic gases~\cite{bloch-2008,greiner-2002,kinoshita-2006,
hofferberth-2007,trotzky-2012,gring-2012,cheneau-2012,
schneider-2012,kunhert-2013,langen-2013,meinert-2013,fukuhara-2013,
ronzheimer-2013,braun-2014,langen-2015}. The paradigm experiment is the so-called global 
{\it quantum quench}~\cite{polkovnikov-2011}, in which a system is initially 
prepared in an eigenstate $|\Psi_0\rangle$ of a many-body Hamiltonian ${\mathcal 
H}$. Then a global parameter of ${\mathcal H}$ is suddenly changed, and the 
system evolves unitarily under the new Hamiltonian ${\mathcal H}'$. At long times 
after the quench the system reaches a steady state, as it has been confirmed by 
experiments~\cite{kinoshita-2006}. In integrable models the 
presence of non-trivial {\it local} conserved quantities, besides the energy, 
strongly affects the dynamics and the nature of the steady state. As for now,  
despite the tremendous theoretical effort~\cite{calabrese-2006,rigol-2007,calabrese-2007,
kollath-2007,manmana-2007,rigol-2008,cramer-2008,barthel-2008,kollar-2008,moeckel-2008,
iucci-2009,rossini-2009,barmettler-2009,biroli-2010,rossini-2010,fioretto-2010,
gogolin-2011,banuls-2011,calabrese-2011,rigol-2011,calabrese-2012,caux-2012,essler-2012,
cazalilla-2012,mossel-2012a,collura-2013,mussardo-2013,caux-2013,fagotti-2013a,fagotti-2013,
sotiriadis-2014,fagotti-2014,essler-2014,goldstein-2014,de-nardis-2014,wright-2014,
pozsgay-2014,wouters-2014,mestyian-2015,ilievski-2015a}, 
it is still unclear whether such steady-state can be described by a statistical 
ensemble, and how to construct it. 

It has been proposed that the long-time stationary value of a generic 
local operator ${\mathcal O}$ is described by a Generalized Gibbs Ensemble~\cite{
rigol-2007,rigol-2008} (GGE) as $\langle{\mathcal O}\rangle\equiv\textrm{Tr}({\mathcal O}
\rho^{GGE})$. Here $\rho^{GGE}$ extends the Gibbs density matrix by including  all 
the extra conserved quantities $ {\mathcal I}_j$ (charges) as 
\begin{equation}
\rho^{GGE}=Z^{-1}\exp\big(-\lambda_j{\mathcal I}_j\big). 
\label{rho-gge}
\end{equation}
In~\eqref{rho-gge}, and in the rest of the paper, repeated indices are summed over. 
$Z$ is a normalization factor. The $\lambda_j$ are Lagrange multipliers to be fixed 
by imposing $\langle\Psi_0|{\mathcal I}_j|\Psi_0\rangle=\langle{\mathcal 
I}_j\rangle$, and ${\mathcal I}_2={\mathcal H}'$ is the post-quench Hamiltonian. 
In realistic situations one deals with the truncated GGE~\cite{fagotti-2013} (TGGE), 
i.e., considering only the ``most local'' charges. 

While the validity of the GGE has been largely confirmed in non-interacting 
theories~\cite{calabrese-2011,calabrese-2012,fagotti-2013,kcc14,kcc14a}, in interacting ones the 
scenario is far less clear (see Ref.~\onlinecite{fagotti-2014} for numerical results in an 
interacting spin chain). For Bethe ansatz solvable models the so-called Quench Action 
method~\cite{caux-2013} allows for an exact treatment of the steady state, 
provided that the overlap between the initial state $|\Psi_0\rangle$ and the eigenstates 
of ${\mathcal H}'$ are known. In several cases the Quench Action is in disagreement with 
the TGGE~\cite{de-nardis-2014,pozsgay-2014,wouters-2014,mestyian-2015}, whereas it is 
supported by numerical simulations~\cite{pozsgay-2014}. 
The origin of this discrepancy remained unknown until very recently. In 
Ref.~\onlinecite{ilievski-2015a} it has been shown that it is possible to 
``repair''  the GGE by including the quasi-local charges~\cite{prosen-2014,
pereira-2014,ilievski-2015}. Remarkably, this repaired GGE is in 
perfect agreement with the Quench Action~\cite{ilievski-2015a}, confirming that 
the description of the steady state with the GGE is correct, provided that the 
appropriate set of local and quasi-local charges is considered. 

%%%%%%%%%%%%%%%%%%%%%%%%%%%%%%%%%e
\begin{figure*}[t]
\includegraphics*[width=0.93\linewidth]{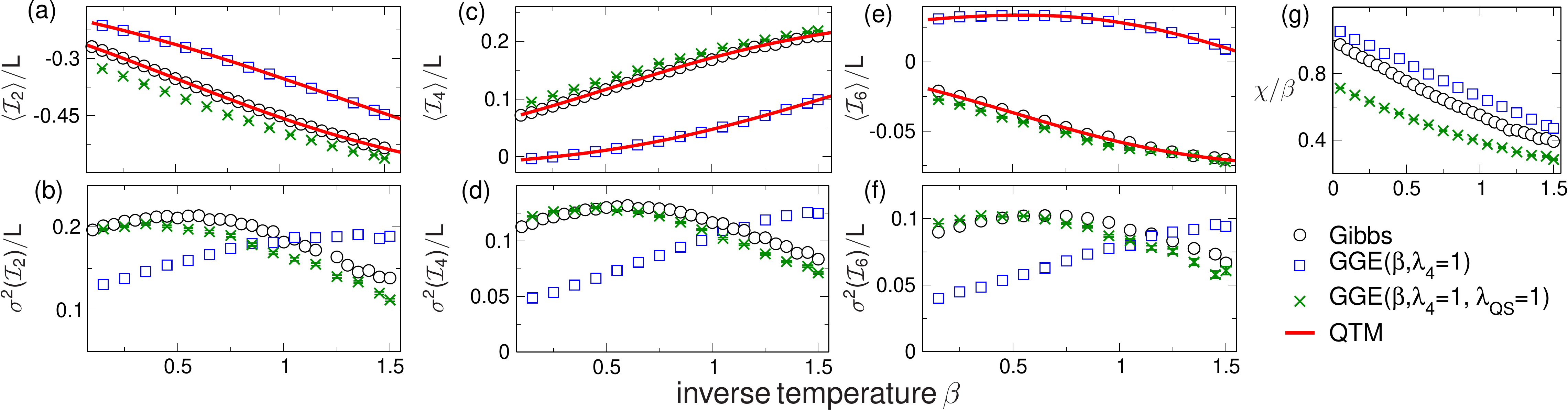}
\caption{The Generalized Gibbs Ensemble (GGE) for the Heisenberg spin chain with 
 $L=16$ sites: Numerical results obtained using the Hilbert space Monte 
 Carlo sampling approach. Only the first two even conserved charges ${\mathcal I}_2,{\mathcal I}_4$ 
 and the first quasi-local one ${\mathcal I}_{QS}$ are included in the GGE. 
 ${\mathcal I}_2$ is the Hamiltonian. In all 
 the panels different symbols correspond to different  values of the Lagrange multipliers 
 $\lambda_4,\lambda_{QS}$. The circles correspond to the Gibbs ensemble, i.e., 
 $\lambda_4=\lambda_{QS}=0$. The $x$-axis shows the inverse temperature  $\lambda_2=\beta$. 
 (a) The GGE average $\langle {\mathcal I}_2/L\rangle$. (b) Variance 
 of the GGE fluctuations $\sigma^2({\mathcal I}_2)/L\equiv(\langle {\mathcal I}_2^2
 \rangle-\langle {\mathcal I}_2\rangle^2)/L$ as a function of $\beta$. (c)(d) and (e)(f): 
 Same as in (a)(b) for ${\mathcal I}_4$ and ${\mathcal I}_6$, respectively. In all panels 
 the  lines  are the Quantum Transfer Matrix (QTM) results. 
 (g) $\chi/\beta$ plotted versus $\beta$, with 
 $\chi$ being the magnetic susceptibility per site. 
}
\label{fig1}
\end{figure*}
%%%%%%%%%%%%%%%%%%%%%%%%%%%%%%%%%%

On the other hand, numerical methods, such as the time dependent density 
matrix renormalization group~\cite{white-2004,daley-2004} (tDMRG), have been mostly 
used to simulate the post-quench dynamics in microscopic models. However, no numerical 
attempt to explore the GGE {\it per se} has been undertaken yet. The aim of this work is to 
provide a Monte-Carlo-based framework for studying the GGE, and its possible extensions,   
in Bethe ansatz solvable models. We restrict ourselves to finite-size systems.  
Thermodynamic quantities can be extracted by a standard finite-size scaling analysis. 
Moreover, as finite-size corrections decay exponentially with system size~\cite{iyer-2015}, 
moderately large systems are sufficient to access the thermodynamic limit. 
The method relies on the detailed knowledge of the Hilbert space structure provided by 
the Bethe ansatz formalism, and on the Bethe-Gaudin-Takahashi (BGT) equations~\cite{takahashi-1971,
taka-book}. The key idea is to sample the model Hilbert space according to 
the GGE probability measure given in~\eqref{rho-gge}. We should mention that the same 
idea has been already explored in Ref.~\onlinecite{gu-2005} for the Gibbs ensemble.
The method allows one to obtain GGE expectation value for generic observables, provided 
that their expression in terms of the roots of the BGT equations are known. Remarkably, 
it is also possible to extract the steady-state roots distributions, which encode 
the complete information about the (GGE) ensemble. It is also straightforward to extend 
the GGE including in~\eqref{rho-gge} arbitrary functions of the BGT roots. This could 
be useful, for instance, to investigate the effects of quasi-local charges. Finally, 
we should mention that, in principle, GGE averages of local observables can be computed 
using exact diagonalization or Quantum Monte Carlo. However, both these methods require 
the operatorial expression of the conserved charges (see Ref.~\onlinecite{grabowski-1995} 
for the $XXX$ chain), whereas our results rely only on their expression (typically simple) 
in terms of the BGT roots. 

We benchmark the approach focusing on the spin-$\frac{1}{2}$ isotropic 
Heisenberg chain ($XXX$ chain), which is the venerable prototype of integrable 
models~\cite{bethe-1931}. We consider several TGGEs 
(cf.~\eqref{rho-gge}) constructed including ${\mathcal I}_2,{\mathcal I}_4$, and the 
first of the recently discovered~\cite{ilievski-2015,ilievski-2015a} 
quasi-local charges ${\mathcal I}_{QS}$ ($H_1^2$ in Ref.~\onlinecite{ilievski-2015a}). 
We focus on the conserved charges averages $\langle{\mathcal I}_j/L\rangle$, 
and on their ensemble fluctuations $\sigma^2({\mathcal I}_j)\equiv\langle{\mathcal I}_j^2
\rangle-\langle{\mathcal I}_j\rangle^2$, which  are related to well-known physical 
observables, such as the energy density, and the specific heat. We also compute the spin susceptibility 
per site $\chi$. Already for a chain with $L=16$ sites the Monte Carlo data perfectly agree 
with both the standard Thermodynamic Bethe Ansatz~\cite{mossel-2012} (TBA) and the Quantum Transfer 
Matrix approach~\cite{fagotti-2013a,pozsgay-2013} (QTM). Notice that this is the first direct 
numerical verification of the QTM approach in the $XXX$ chain. 
Finally, we extract the BGT roots distributions for both the Gibbs ensemble and the GGE. 
In both cases the finite-size effects are negligible for small roots, which are the 
relevant ones to describe the long-wavelength physics. For the Gibbs ensemble we compare 
our numerical data with standard finite-temperature Thermodynamic Bethe Ansatz (TBA) results, 
finding excellent agreement.

%#############-- THE HEISENBERG SPIN CHAIN --########################################
\paragraph*{The-Heisenberg-spin-chain.---}

The $XXX$ chain with $L$ sites is defined by the Hamiltonian 
\begin{align}
\label{xxx-ham}
{\mathcal H}\equiv J\sum\limits_{i=1}^L\left[\frac{1}{2}(S_i^+S^-_{i+1} 
+S_i^{-}S_{i+1}^+)+S_i^zS_{i+1}^z-\frac{1}{4}\right],  
\end{align}
where $S^{\pm}_i\equiv (\sigma_i^x\pm i\sigma_i^y)/2$ are spin operators acting on the 
site $i$, $S_i^z\equiv\sigma_i^z/2$, and $\sigma^{x,y,z}_i$ the Pauli matrices. We fix 
$J=1$ and use periodic boundary conditions, identifying sites $L+1$ and $1$. The total 
magnetization $S_{T}^z\equiv\sum_iS_i^z=L/2-M$, with $M$ number of down spins (particles), 
commutes with~\eqref{xxx-ham}, and it is here used to label its eigenstates. 

In the Bethe ansatz formalism each eigenstate of~\eqref{xxx-ham} is univocally identified 
by $M$ parameters  $\{x_\alpha\in\mathbb{C}\}_{\alpha=1}^M$. In the limit $L\to\infty$ they 
form ``string'' patterns along the imaginary axis of the complex 
plane (string hypothesis~\cite{bethe-1931,taka-book}). Strings of length $1\le n\le M$ 
(so-called $n$-strings) are parametrized as $x_{n;\gamma}^{j}=x_{n;\gamma}-i(n-1-2j)$. 
Here $x_{n;\gamma}\in\mathbb{R}$ is the string real part (string center), $j=0,1,\dots,n-1$ 
labels different string components, and $\gamma$ denotes different string centers. The 
string hypothesis is not correct for finite chains, although deviations typically decay 
exponentially with $L$. Physically, the $n$-strings correspond to eigenstate components 
containing $n$-particle bound states. The $\{x_{n;\gamma}\}$ are obtained as the roots 
of the Bethe-Gaudin-Takahashi (BGT) equations~\cite{takahashi-1971,taka-book} 
\begin{equation}
L\vartheta_n(x_{n;\gamma})=2\pi I_{n;\gamma}+\sum\limits_{(m,\beta)
\ne(n,\gamma)}\Theta_{m,n}(x_{n;\gamma}-x_{m;\beta}).
\label{bt-eq}
\end{equation}
Here $\vartheta_n(x)\equiv2\arctan(x/n)$, $\Theta_{m,n}(x)$ is the scattering 
phase between different roots~\cite{taka-book}, and $I_{n;\gamma}\in\frac{1}{2}\mathbb{Z}$ are 
the so-called Bethe-Gaudin-Takahashi quantum numbers. The $I_{n;\gamma}$ satisfy the upper 
bound $|I_{n;\gamma}|\le I_{\textrm{MAX}}(n,L,M)$, with $I_{\textrm{MAX}}$ a known 
function~\cite{taka-book} of $n,L,M$. 
Every choice of $I_{n;\gamma}$ identifies an eigenstate of~\eqref{xxx-ham}. 
We define the ``string content'' of each eigenstate as ${\mathcal S}\equiv\{s_1,\dots,
s_M\}$, with $0\le s_n\le \lfloor M/n\rfloor$ the number of $n$-strings.  
The local conserved charges ${\mathcal I}_j$ of the $XXX$ chain are given as   
\begin{equation}
\label{I-def}
\left.{\mathcal I}_{j+1}\equiv\frac{i}{(j-1)!}\frac{d^j}{dy^j}\log\Lambda
(y)\right|_{y=i}, 
\end{equation}
where $\Lambda(y)$ is the eigenvalue of the quantum transfer matrix~\cite{kor-book}, 
with $y$ a spectral parameter. ${\mathcal I}_2$ is the $XXX$ Hamiltonian. 
The analytic expression of ${\mathcal I}_j$ in terms of the Pauli matrices is known~\cite{grabowski-1995}  
for $j\le 10$. The support of ${\mathcal I}_j$, i.e., the number of 
adjacent sites where ${\mathcal I}_j$ acts non trivially, increases linearly with $j$, i.e., 
larger $j$ correspond to less local charges.  
The eigenvalues of ${\mathcal I}_j$ on a generic eigenstate are 
obtained by summing the contributions of the different BGT roots {\it independently}. 
For instance, the energy eigenvalue is obtained as $E=2\sum_{n,\gamma} n/(n^2+x^2_{n;\gamma})$. 
A similar result holds true for the quasi-local charges~\cite{ilievski-2015a}.

%#############-- HILBERT SPACE MONTE CARLO --########################################
\paragraph*{The-Hilbert-space-Monte-Carlo-sampling.---}

For a finite chain the GGE (cf.~\eqref{rho-gge}) can be obtained by importance 
sampling~\cite{landau-binder} of the eigenstates of~\eqref{xxx-ham}. One starts with an 
initial $M$-particle eigenstate, with string content ${\mathcal S}=\{s_1,\dots,s_M\}$, and 
identified by a BGT quantum number configuration ${\mathcal C}=\{I_{n;\gamma}\}_{n=1}^M$ 
($\gamma=1,\dots,s_n$). The corresponding charges eigenvalues are $\{{\mathcal I_j}\}$. 
Then a new eigenstate is generated with a Monte Carlo scheme. 
Each Monte Carlo step (mcs) consists of three moves:
\begin{enumerate}
\item[\circled{1}] Choose a new particle number sector $M'$, and string content ${\mathcal S}'
 \equiv\{s_1',\dots,s_{M'}'\}$ with probability~\cite{faddeev-1996} ${\mathcal P}(M',{\mathcal S}')$
\vspace{-5pt}
\begin{equation}
\label{dos}
{\mathcal P}(M',{\mathcal S}')=\frac{1}{B(L,L/2)}\prod_{i=1}^{M'} B\left({\mathcal L}_i,{
\mathcal S}'_i\right).
\end{equation}
\vspace{-16pt}
\item[\circled{2}] Generate a new quantum number configuration ${\mathcal C}'$ compatible with 
 the ${\mathcal  S}'$ obtained in step $1$. Solve the corresponding BGT 
 equations~\eqref{bt-eq}. 
\item[\circled{3}] Calculate the charge eigenvalues ${\mathcal I}_j'$ and accept the new 
eigenstate with the Metropolis probability:
\begin{equation}
\label{metropolis}
\textrm{Min}\Big\{1,\frac{L-2M'+1}{L-2M+1}e^{-\lambda_j({\mathcal I}'_j-
{\mathcal I}^{}_j)}\Big\}.
\end{equation}
\end{enumerate}
In~\eqref{dos} $B(x,y)\equiv x!/(y!(x-y)!)$ is the Newton binomial and ${\mathcal L}_i\equiv L-\sum_{j=1}^{M'}t_{ij}
{\mathcal S}'_j$, with $t_{ij}\equiv2\textrm{Min}(i,j)-\delta_{ij}$.
In~\eqref{metropolis} the factor in front of the exponential takes into account 
that ${\mathcal I}_j$ and the observables that we consider are invariant under $SU(2)$ rotations. 
Crucially, the steps $1$ and $2$ are necessary to account for the density of states of the model 
(equivalently, the Yang-Yang entropy, see below), 
and are the same as for the Gibbs ensemble~\cite{gu-2005}. The iteration of $1$-$3$ defines a Markov 
chain, which, after some thermalization steps, generates eigenstates distributed according 
to~\eqref{rho-gge}. Interestingly, by trivially modifying~\eqref{metropolis} it is possible to 
simulate more exotic ensembles in which, in addition to ${\mathcal I}_j$, one considers 
arbitrary functions of the BGT roots.  
%
%%%%%%%%%%%%%%%%%%%%%%%%%%%%%%%%%e
\begin{figure}[t]
\includegraphics*[width=0.93\linewidth]{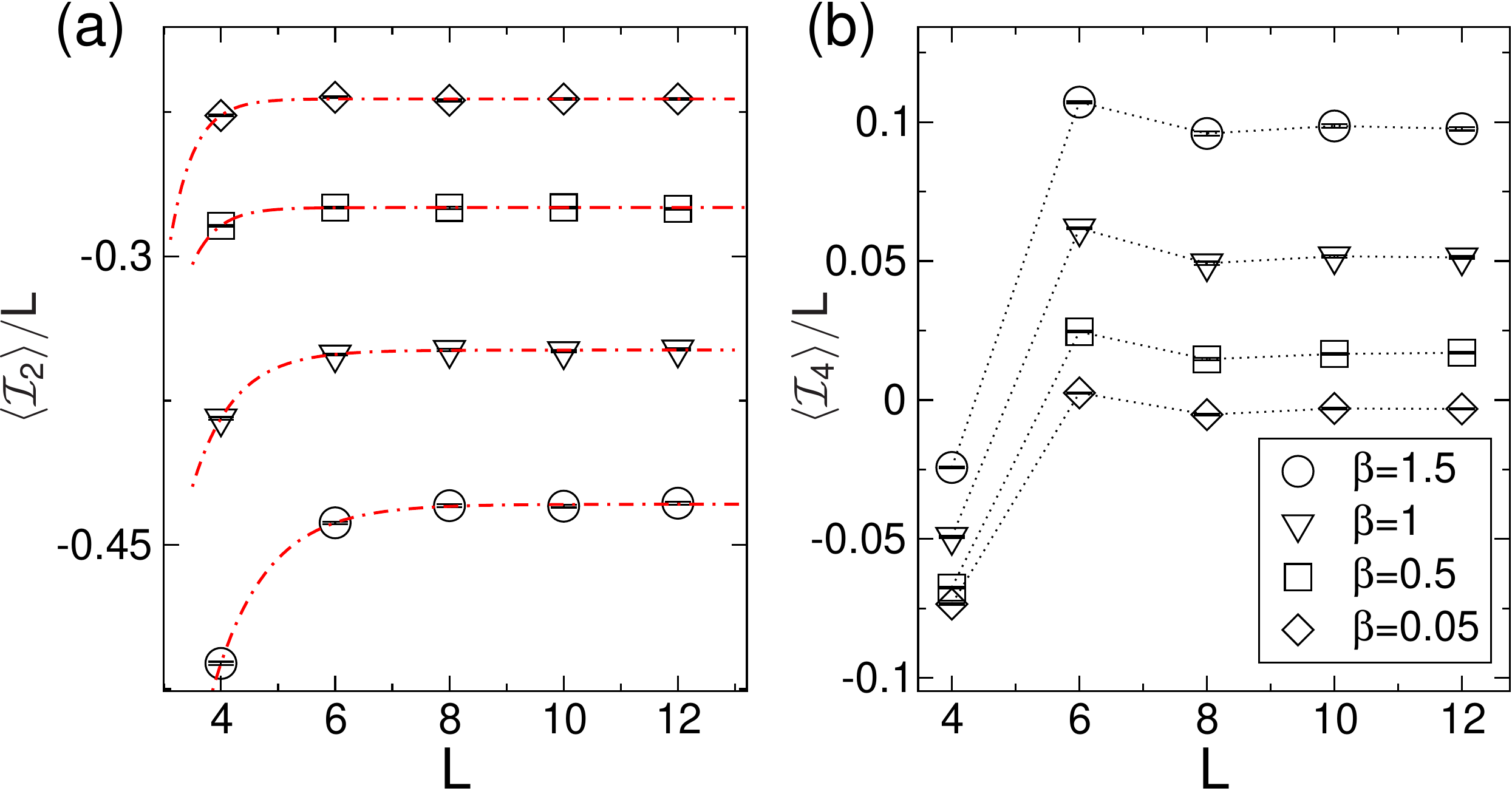}
\caption{Finite-size scaling of the GGE averages in the Heisenberg chain: Numerical results 
 obtained from the Hilbert space Monte Carlo sampling. Here the GGE is constructed including 
 ${\mathcal I}_2,{\mathcal I}_4$, with Lagrange multipliers $\lambda_2=\beta,
 \lambda_4=1$. (a) $\langle {\mathcal I}_2/L\rangle$ plotted versus the chain size 
 $L$ for several values of $\beta$. The dash-dotted lines are exponential fits. (b) Same as 
 in (a) for ${\mathcal I}_4$.
}
\label{fig2}
\end{figure}
%%%%%%%%%%%%%%%%%%%%%%%%%%%%%%%%%%
%
The GGE average $\langle{\mathcal O}\rangle$ of a generic operator is obtained as  
\begin{equation}
\label{gge-mc}
\langle{\mathcal O}\rangle=\lim\limits_{N_\textrm{mcs}\to\infty}\frac{1}
{N_\textrm{mcs}}\sum\limits_{|s\rangle}\langle s|{\mathcal O}
|s\rangle,
\end{equation}
where $N_{\textrm{mcs}}$ is the total number of eigenstates $|s\rangle$ sampled in 
the Monte Carlo. Moreover, for all the observables considered here the contributions 
of the BGT roots can be summed independently, i.e., 
\begin{equation}
\label{gge-mc-1}
\langle s|{\mathcal O}|s\rangle=\sum\limits_{n,\gamma}f_{{\mathcal O}}(x_{n;\gamma}), 
\end{equation}
where $x_{n;\gamma}$ are the roots identifying the eigenstate $|s\rangle$, and 
$f_{{\mathcal O}}(x)$ depends on the observable.

%#############-- GGE FOR LOCAL OBSERVABLES --########################################
\paragraph*{The-GGE-for-local-observables.---} 

The correctness of the Monte Carlo approach is illustrated in Fig.~\ref{fig1}, considering 
the charge densities $\langle {\mathcal I}_j/L\rangle$ (panels (a)(c)(e) in the Figure), 
and the variance of their ensemble fluctuations $\sigma^2({\mathcal I}_j)/L\equiv
(\langle{\mathcal I}_j^2\rangle-\langle{\mathcal I}_j\rangle^2)/L$ (panels (b)(d)(f)). 
Panel (g) plots $\chi/\beta$, with $\chi$ the spin susceptibility. Notice that 
$\langle{\mathcal I}_2/L\rangle$ is the energy 
density, while $\sigma^2({\mathcal I}_2)/L$ is related to the specific heat. In all 
panels the data correspond to the TGGE constructed 
with the first two even charges ${\mathcal I}_2,{\mathcal I}_4$, and the first 
non-trivial quasi-local charge ${\mathcal I}_{QS}$~\cite{ilievski-2015,ilievski-2015a}. 
Different symbols correspond to different values of the associated Lagrange multipliers, 
namely $\lambda_4=\lambda_{QS}=0$ (Gibbs ensemble, circles in the Figure), $\lambda_4=1$ and 
$\lambda_{QS}=0$ (squares), and $\lambda_4=0,\lambda_{QS}=1$ (crosses). 
In all panels the $x$-axis shows the inverse temperature $\lambda_2=\beta$. The data 
are Monte Carlo averages with $N_{\textrm{mcs}}=5\cdot 10^5$ (cf.~\eqref{gge-mc}). 
As expected, the different ensembles give different expectation values, implying that 
local observables are able to distinguish different GGEs.
In Fig.~\ref{fig1} the continuous lines are the analytic results obtained 
in the thermodynamic limit using the QTM approach. These fully match the Monte 
Carlo data, signaling  that finite-size effects are negligible already for $L=16$.  

%%%%%%%%%%%%%%%%%%%%%%%%%%%%%%%%%e
\begin{figure*}[t]
\includegraphics*[width=0.99\linewidth]{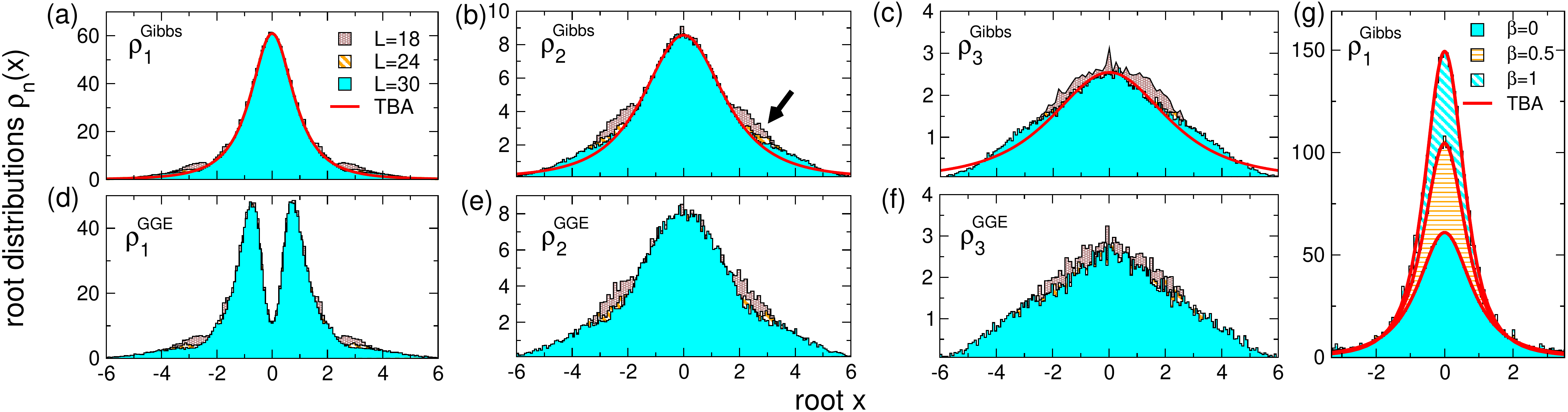}
\caption{The root distributions $\rho_n(x)$ (for $n=1,2,3$) for the infinite temperature 
 Gibbs (panels (a)-(c)) and the GGE equilibrium states (panels (d)-(f)): Numerical  
 results for the Heisenberg spin chain obtained using the Hilbert space Monte Carlo 
 sampling. Here the GGE is constructed including only ${\mathcal I}_2$ and ${\mathcal I}_4$ 
 with fixed Lagrange multipliers $\lambda_2=0$ and $\lambda_4=1$. In all the panels the 
 data are the histograms of the $n$-strings roots sampled in the Monte Carlo.
 The width of the histogram bins is $\Delta x=2/L$, with $L$ the chain size. 
 In each panel different histograms correspond to different $L$. All the data are divided 
 by $10^3$ for convenience. In (b) the arrow highlights the finite-size effects. In 
 (a)-(c) the lines are the Thermodynamic Bethe Ansatz (TBA) results. (g) Finite-temperature 
 effects: Monte Carlo data for $\rho^{\textrm{Gibbs}}_1$ for different values of the 
 inverse temperature $\beta$.
}
\label{fig3}
\end{figure*}
%%%%%%%%%%%%%%%%%%%%%%%%%%%%%%%%%%

%#############-- FINITE-SIZE CORRECTIONS --########################################

The finite-size corrections are more carefully investigated in Fig.~\ref{fig2},   
plotting $\langle{\mathcal I}_2\rangle$ and $\langle {\mathcal I}_4
\rangle$ (panels (a) and (b), respectively) versus $\beta$. We focus on 
the TGGE with $\lambda_2=\lambda_{QS}=0$ and $\lambda_4=1$. Clearly,  
finite-size effects decay exponentially~\cite{iyer-2015} with $L$ for any $\beta$. In (a) the dashed 
lines are fits to $c_1+c_2\exp(-c_3L)$, with $c_1,c_2,c_3$ fitting parameters. 
Finite-size  corrections are larger at lower temperature, and increase with the 
range of the operator (compare panels  (a) and (b) in Fig.~\ref{fig2}), as expected. 

%#############-- EXTRACTING THE ROOT DENSITIES --###################################
\paragraph*{Extracting-the-root-distributions.---}

In the thermodynamic limit in each $n$-string sector the roots of~\eqref{bt-eq} become 
dense. Thus, instead of the eigenstates, one considers the corresponding root 
distributions $\pmb{\rho}\equiv\{\rho_n\}_{n=1}^\infty$. Formally, $\rho_n\equiv
\lim_{L\to\infty}[L(x_{n;\gamma+1}-x_{n;\gamma})]^{-1}$. The GGE average of a 
generic observable ${\mathcal O}$ becomes a functional integral as~\cite{yang-1969,mossel-2012} 
\begin{equation}
\label{th-limit}
\textrm{Tr}\big\{\exp\big({\lambda_j{\mathcal I}_j}\big){\mathcal O}\big\}
\rightarrow\int{\mathcal D}\pmb{\rho}\exp\big(S[\pmb{\rho}]+
\lambda_j{\mathcal I}_j[\pmb{\rho}]\big){\mathcal O}[\pmb{\rho}].
\end{equation}
Here $S[\pmb{\rho}]$ is the Yang-Yang entropy, which counts the number of eigenstates 
leading to the same $\pmb{\rho}$ in the thermodynamic limit, and it is extensive. 
In~\eqref{th-limit} it is assumed that ${\mathcal O}$ becomes a smooth functional 
of $\pmb{\rho}$ in the thermodynamic limit. Eq.~\eqref{gge-mc-1} becomes 
\begin{equation}
\label{th-limit-1}
\langle s|{\mathcal O}|s\rangle\to\sum\limits_{n}\int dx\rho_n(x)f_{{\mathcal O}}(x).
\end{equation}
Since both $S[\pmb{\rho}]$ and ${\mathcal I}_j[\pmb{\rho}]$ are extensive, the functional 
integral in~\eqref{th-limit} is dominated by the saddle point~\cite{mossel-2012} 
$\pmb{\rho}^{sp}$, with $\delta(S+\lambda_j{\mathcal I}_j)/\delta\pmb{\rho}|_{\pmb{\rho}=
\pmb{\rho}^{sp}}=0$. Here $\pmb{\rho}^{sp}$ acts as a representative 
state for the ensemble, and it contains the full information about the GGE equilibrium steady 
state. Eq.~\eqref{gge-mc} and~\eqref{th-limit-1} imply that in the thermodynamic limit 
the histograms of the BGT roots sampled in the Monte Carlo converge to $\pmb{\rho}^{sp}$.  

This is numerically supported in Fig.~\ref{fig3}. Panels (a)-(c) plot the  
root distributions $\rho^{sp}_n(x)$ for $n=1,2,3$ as a function of $x$ for the representative 
state (saddle point) of the infinite-temperature Gibbs ensemble. In each panel 
the different histograms correspond to different chain sizes $18\le L\le 30$. The data 
are obtained using $5\cdot 10^5$ Monte Carlo steps. The width of the histogram bins is 
varied with $L$ as $2/L$. In all the panels the full lines are the analytical   
Thermodynamic Bethe Ansatz~\cite{taka-book} (TBA) results. Clearly, deviations from 
the TBA vanish upon increasing the chain size (see for instance the arrow in panel (b)). 
Moreover, the corrections are larger on the tails of the distributions. This is expected 
since large roots correspond to large quasi-momenta, which are more sensitive to the 
lattice effects. Finally, finite-size effects increase with $n$, i.e., with the bound 
state sizes, as expected. The results for the finite-temperature Gibbs ensemble 
are reported in Fig.~\ref{fig1} (g), for $\beta=1/2$ and $\beta=1$ (the different 
histograms). We focus on $\rho^{sp}_1(x)$, restricting ourselves to $L=30$. The 
infinite temperature histogram is reported for comparison. The continuous lines are now 
finite-temperature TBA results, and perfectly agree with the Monte Carlo data. 
Upon lowering the temperature the height of the peak at $x=0$ increases. This reflects 
that at $\beta=\infty$ the tail of the root distributions vanish exponentially, whereas 
for $\beta=0$ they are~\cite{taka-book} $\sim 1/x^4$. 
Finally, panels (d)-(f) plot $\rho_n(x)$ for the TGGE constructed with ${\mathcal I}_2,
{\mathcal I}_4$ at fixed  $\lambda_2=0$, $\lambda_4=1$ and for $L=30$. Interestingly, 
in contrast with the thermal case (see (a)), $\rho^{sp}_1$ exhibits a double peak 
at small $x$. Similar to the infinite-temperature Gibbs ensemble ((a)-(c) in 
the Figure), the data suggest that for $L=30$ finite-size effects are negligible, 
at least for $-2\le x\le 2$.

%#########-- CONCLUSIONS --################################################
\paragraph*{Conclusions.---}

We presented a Monte-Carlo-based scheme for simulating the truncated Generalized 
Gibbs ensemble (TGGE) in finite-size integrable models. The key idea is the importance 
sampling of the model eigenstates using the GGE probability measure. The method relies 
on the Bethe ansatz formalism, and, in particular, on the Bethe-Gaudin-Takahashi (BGT) 
equations. The thermodynamic limit can be accessed by standard finite-size scaling analysis. 
For local quantities we observed that the finite-size corrections decay exponentially 
with the system size. 
Remarkably, the method allows to extract in a simple way the steady-state BGT root 
distributions, which contain full information about the (GGE) ensemble averages 
in the thermodynamic limit. Finally, it is possible to simulate extensions of 
the GGE, in which, besides the integral of motion, one includes arbitrary functions 
of the BGT roots. We benchmarked the method focusing on the spin-$\frac{1}{2}$ isotropic 
Heisenberg chain. Specifically, we compared the Monte Carlo results with the standard  
Thermodynamic Bethe ansatz and the Quantum Transfer Matrix approach, finding excellent 
agreement. Finally, we simulated an extended GGE obtained by including the first non-trivial 
quasi-local charge. 

As an interesting research direction, we mention that it would be useful to generalize the 
method to simulate the GGE at fixed value of the conserved charges. This should be possible 
using the standard microcanonical Monte Carlo techniques that have been developed in lattice 
gauge theory~\cite{creutz-1983} and in molecular dynamics simulations~\cite{lustig-1998}. 
Finally, by including in~\eqref{metropolis} the overlap contribution 
$\log|\langle\Psi_0|\Psi_j\rangle|$, with $|\Psi_0\rangle$ the pre-quench initial state, 
and $|\Psi_j\rangle$ the eigenstates of the model,  it should be possible to simulate the 
Quench Action~\cite{prep}.

\paragraph*{Acknowledgements.---}
I would like to thank Maurizio Fagotti for providing the analytical Quantum Transfer Matrix 
results in Fig.~\ref{fig1} and Lorenzo Piroli for the finite-temperature TBA in Fig.~\ref{fig3}. 
I would like to thank P.~Calabrese, M.~Fagotti, F.~Essler, and L.~Piroli,  for useful 
discussions and comments. I acknowledge financial support by the ERC under Starting 
Grant 279391 EDEQS. The simulations were done on the main SISSA cluster. Using Mathematica 
the simulation of $L=30$ (see Fig.~\ref{fig3}) took approximately $2$ days on a single core 
of a standard commercial CPU. 

%##########################################################################


\begin{thebibliography}{99}
\bibitem{bloch-2008}
I.~Bloch, J.~Dalibard, and W.~Zwerger, Rev.\ Mod.\ Phys.\ {\bf 80}, 
885 (2008).

\bibitem{greiner-2002}
M.~Greiner, O.~Mandel, T. H\"ansch, and I.~Bloch, Nature (London) 
{\bf 419}, 51 (2002). 

\bibitem{kinoshita-2006}
T.~Kinoshita, T.~Wenger, and D.~S.~Weiss, Nature (London) {\bf 440}, 
900 (2008).

\bibitem{hofferberth-2007}
S.~Hofferberth, I.~Lesanovsky, B.~Fischer, T.~Schumm, and J.~Schiedmayer, 
Nature (London) {\bf 449}, 324 (2007). 

\bibitem{trotzky-2012}
S.~Trotzky, Y.-A.~Chen, A.~Flesch, I.~P.~McCulloch, U.~Schollw\"ock, 
J.~Eisert, and I.~Bloch, Nature Phys.\ {\bf 8}, 325 (2012).

\bibitem{gring-2012}
M.~Gring, M.~Kuhnert, T.~Langen, T.~Kitagawa, B.~Rauer, M.~Schreitl, 
I.~Mazets, D.~A.~Smith, E.~Demler, and J.~Schmiedmayer, Science {\bf 337}, 
6100 (2012).

\bibitem{cheneau-2012}
M.~Cheneau, P.~Barmettler, D.~Poletti, M.~Endres, P.~Schaua, T.~Fukuhara, 
C.~Gross, I.~Bloch, C.~Kollath, and S.~Kuhr, Nature (London) {\bf 481}, 
484 (2012).

\bibitem{schneider-2012}
U.~Schneider, L.~Hackeruller, J.~P.~Ronzheimer, S.~Will, S.~Braun, T.~Best, 
I.~Bloch, E.~Demler, S.~Mandt, D.~Rasch, and A.~Rosch, Nature\ Phys.\ 
{\bf 8}, 213 (2012).

\bibitem{kunhert-2013}
M.~Kuhnert, R.~Geiger, T.~Langen, M.~Gring, B.~Rauer,
T.~Kitagawa, E.~Demler, D.~Adu Smith, and J.~Schmiedmayer, Phys.\ Rev.\ 
Lett.\ {\bf 110}, 090405 (2013).

\bibitem{langen-2013}
T.~Langen, R.~Geiger, M.~Kuhnert, B.~Rauer, and J.~Schmiedmayer, 
Nature\ Phys.\ {\bf 9}, 640 (2013).

\bibitem{meinert-2013}
F.~Meinert, M.~J.~Mark, E.~Kirilov, K.~Lauber, P.~Weinmann, 
A.~J.~Daley, and H.-C.~Nagerl, Phys.\ Rev.\ Lett.\ {\bf 111}, 
053003 (2013).

\bibitem{fukuhara-2013}
T.~Fukuhara, A.~Kantian, M.~Endres, M.~Cheneau, P.~Schaua, S.~Hild, C.~Gross, 
U.~Schollw\"ock, T.~Giamarchi, I.~Bloch, and S.~Kuhr, Nature\ Phys.\ {\bf 9}, 
235 (2013).

\bibitem{ronzheimer-2013}
J.~P.~Ronzheimer, M.~Schreiber, S.~Braun, S.~S.~Hodgman, S.~Langer, I.~P.~McCulloch, 
F. Heidrich-Meisner, I.~Bloch, and U.~Schneider, Phys.\ Rev.\ Lett.\ {\bf 110}, 
205301 (2013).

\bibitem{braun-2014}
S.~Braun, M.~Friesdorf, S.~Hodgman, M.~Schreiber, J.~Ronzheimer, A.~Riera, M.~del Rey, 
I.~Bloch, J.~Eisert, and U.~Schneider, PNAS {\bf 112}, 3641 (2015).

\bibitem{langen-2015}
T.~Langen, S.~Erne, R.~Geiger, B.~Rauer, T.~Schweigier, M.~Kuhnert, W.~Rohringer, 
I.~E.~Mazets, T.~Gasenzer, J.~Schmiedmayer, Science {\bf 348}, 6231 (2015). 


\bibitem{polkovnikov-2011}
A.~Polkovnikov, K.~Sengupta, A~Silva, and M.~Vengalattore, Rev.\ Mod.\ Phys.\ 
{\bf 83}, 863 (2011). 


\bibitem{calabrese-2006}
P.~Calabrese and J.~Cardy, Phys.\ Rev.\ Lett.\ {\bf 96}, 136801 (2006). 

\bibitem{rigol-2007}
M.~Rigol, V.~Dunjko, V.~Yurovsky, and M.~Olshanii, Phys.\ Rev.\ Lett.\ 
{\bf 98}, 050405 (2007). 

\bibitem{calabrese-2007}
P.~Calabrese and J.~Cardy, J.\ Stat.\ Mech.\ (2007) P06008.

\bibitem{kollath-2007}
C.~Kollath, A.~M.~L\"auchli, and E.~Altman, Phys.\ Rev.\ Lett.\ 
{\bf 98}, 180601 (2007).

\bibitem{manmana-2007}
S.~R.~Manmana, S.~Wessel, R.~M.~Noack, and A.~Muramatsu, 
Phys.\ Rev.\ Lett.\ {\bf 98}, 210405 (2007).

\bibitem{rigol-2008}
M.~Rigol, V.~Dunjko, and M.~Olshanii, Nature {\bf 452}, 854 (2008). 

\bibitem{cramer-2008}
M.~Cramer, C.~M.~Dawson, J.~Eisert, and T.~J.~Osborne, Phys.\ Rev.\ 
Lett.\ {\bf 100}, 030602 (2008).

\bibitem{barthel-2008}
T.~Barthel and U.~Schollw\"ock, Phys.\ Rev.\ Lett.\ {\bf 100}, 100601 
(2008). 

\bibitem{kollar-2008}
M.~Kollar and M.~Eckstein, Phys.\ Rev.\ A {\bf 78}, 013626 (2008). 

\bibitem{moeckel-2008}
M.~Moeckel and S.~Kehrein, Phys.\ Rev.\ Lett.\ {\bf 100}, 175702 (2008). 

\bibitem{iucci-2009}
A.~Iucci and M.~A.~Cazalilla, Phys.\ Rev.\ A\ {\bf 80}, 063619 (2009).

\bibitem{rossini-2009}
D.~Rossini, A.~Silva, G.~Mussardo, and G.~E.~Santoro, Phys.\ Rev.\ 
Lett.\ {\bf 102}, 127204 (2009).

\bibitem{barmettler-2009}
P.~Barmettler, M.~Punk, V.~Gritsev, E.~Demler, and E.~Altman, Phys.\ Rev.\ 
Lett.\ {\bf 102}, 130603 (2009).

\bibitem{biroli-2010}
G.~Biroli, C.~Kollath, and A.~M.~L\"auchli, Phys.\ Rev.\ Lett.\ 
{\bf 105}, 250401 (2010). 

\bibitem{rossini-2010}
D.~Rossini, S.~Suzuki, G.~Mussardo, G.~E.~Santoro, and A.~Silva, 
Phys.\ Rev.\ B\ {\bf 82}, 144302 (2010).

\bibitem{fioretto-2010}
D.~Fioretto and G.~Mussardo, New\ J.\ Phys.\ {\bf 12}, 
055015 (2010).

\bibitem{gogolin-2011}
C.~Gogolin, M.~P.~Mueller, and J.~Eisert, Phys.\ Rev.\ Lett.\ 
{\bf 106}, 040401 (2011).

\bibitem{banuls-2011}
M.~C.~Ba\~nuls, J.~I.~Cirac, and M.~B.~Hastings, Phys.\ Rev.\ Lett.\ 
{\bf 106}, 050405 (2011). 

\bibitem{calabrese-2011}
P.~Calabrese, F.~H.~L.~Essler, and M.~Fagotti, Phys.\ Rev.\ Lett.\ {\bf106}, 227203 (2011).

\bibitem{rigol-2011}
M.~Rigol and M.~Fitzpatrick, Phys.\ Rev.\ A {\bf 84}, 033640 (2011).

\bibitem{calabrese-2012}
P.~Calabrese, F.~H.~L.~Essler, and M.~Fagotti, J.\ Stat.\ Mech.\ (2012) P07016.

\bibitem{caux-2012}
J.-S.~Caux and R.~M.~Konik, Phys.\ Rev.\ Lett.\ {\bf 109}, 175301 (2012).

\bibitem{essler-2012}
F.~H.~L.~Essler, S.~Evangelisti, and M.~Fagotti, Phys.\ Rev.\ Lett.\ 
{\bf 109}, 247206 (2012). 

\bibitem{cazalilla-2012}
M.~A.~Cazalilla, A.~Iucci, and M.-C.~Chung, Phys.\ Rev.\ E {\bf 85}, 
011133 (2012). 

\bibitem{mossel-2012a}
J.~Mossel and J.-S.~Caux, New\ J.\ Phys.\ {\bf 14} 075006 (2012).

\bibitem{collura-2013}
M.~Collura, S.~Sotiriadis and P.~Calabrese, Phys.\ Rev.\ Lett.\ {\bf 110}, 245301 (2013)

\bibitem{mussardo-2013}
G.~Mussardo, Phys.\ Rev.\ Lett.\ {\bf 111}, 100401 (2013).

\bibitem{caux-2013}
J.-S.~Caux and F.~H.~L.~Essler, Phys.\ Rev.\ Lett.\ {\bf 110}, 
257203 (2013). 

\bibitem{fagotti-2013a}
M.~Fagotti and F.~H.~L.~Essler, J.\ Stat.\ Mech.\ (2013), P07012. 

\bibitem{fagotti-2013}
M.~Fagotti and F.~H.~L.~Essler, Phys.\ Rev.\ B\ {\bf87}, 245107 (2013).

\bibitem{sotiriadis-2014}
S.~Sotiriadis and P.~Calabrese, J.\ Stat.\ Mech.\ (2014) P07024. 

\bibitem{fagotti-2014}
M.~Fagotti, M.~Collura, F.~H.~L.~Essler, and P.~Calabrese, Phys.\ Rev.\ B {\bf 89}, 
125101 (2014).

\bibitem{essler-2014}
F.~H.~L.~Essler, S.~Kehrein, S.~R.~Manmana, and N.~J.~Robinson, Phys.\ Rev.\ B {\bf 89}, 
165104 (2014).

\bibitem{goldstein-2014}
G.~Goldstein and N.~Andrei, arXiv:1405.4224. 


\bibitem{de-nardis-2014}
J.~De Nardis, B.~Wouters, M.~Brockmann, and J.-S.~Caux, Phys.\ Rev.\ A {\bf 89}, 
033601 (2014). 

\bibitem{wright-2014}
T.~M.~Wright, M.~Rigol, M.~J.~Davis, and K.~V.~Kheruntsyan, Phys.\ Rev.\ Lett.\ {\bf 113}, 
050601 (2014).

\bibitem{pozsgay-2014}
B.~Pozsgay, M.~Mesty\'an, M.~A.~Werner, M.~Kormos, G.~Zar\'and, and G.~Tak\'acs,
Phys.\ Rev.\ Lett.\ {\bf 113}, 117203 (2014). 

\bibitem{wouters-2014}
B.~Wouters, M.~Brockmann, J.~De~Nardis, D.~Fioretto, M.~Rigol, and J.-S.~Caux, 
Phys.\ Rev.\ Lett.\ {\bf 113}, 117202 (2014). 

\bibitem{mestyian-2015}
M~Mesty\'an, B.~Pozsgay, G.~Tak\'acs, and M.~A.~Werner, J.\ Stat.\ Mech.\ (2015) 
P04001.


\bibitem{ilievski-2015a}
E.~Ilieveski, J.~De~Nardis, B.~Wouters, J.-S.~Caux, F.~H.~Essler, and T.~Prosen, 
arXiv:1507.02993. 

\bibitem{kcc14}
M.~Kormos, M.~Collura, and P.~Calabrese, Phys.\ Rev.\ A {\bf 89}, 013609 
(2014).


\bibitem{kcc14a}
P.~P.~Mazza, M.\ Collura, M.\ Kormos, and P.\ Calabrese, J.\ Stat.\ Mech.\  
(2014) P11016.

\bibitem{prosen-2014}
T.~Prosen, Nucl.\ Phys.\ B\ {\bf 886}, (2014) 1177.

\bibitem{pereira-2014}
R.~G.~Pereira, V.~Pasquier, J.~Sirker, and I.~Affleck, J.\ Stat.\ Mech.\ 
(2014) P09037. 

\bibitem{ilievski-2015}
E.~Ilievski, M.~Medejak, and T.~Prosen, arXiv:1506.05049.

\bibitem{white-2004}
S.~R.~White and A.~E.~Feiguin, Phys.\ Rev.\ Lett.\ {\bf 93}, 076401 (2004).

\bibitem{daley-2004}
A.~J.~Daley, C.~Kollath, U.~Schollock, and G.~Vidal, J.\ Stat.\ Mech.\ (2004) P04005.

\bibitem{iyer-2015}
D.~Iyer, M.~Srednicki, and M.~Rigol, Phys.\ Rev.\ E\ {\bf 91}, 062142 (2015).

\bibitem{takahashi-1971} 
M.~Takahashi, Prog.\ Theor.\ Phys.\ {\bf 46}, 401 (1971). 

\bibitem{taka-book}
M.~Takahashi, {\it Thermodynamics of one-dimensional solvable models}, 
Cambridge University Press, Cambridge, 1999. 

\bibitem{gu-2005}
S.-J.~Gu, N.~M.~R.~Peres, Y.-Q.~Li, Eur.\ Phys.\ J.\ B {\bf 48}, 157 (2005). 

\bibitem{grabowski-1995}
M.~P.~Grabowski and P.~Mathieu, Ann.\ Phys.\ N.Y. {\bf 243}, 
299 (1995). 

\bibitem{bethe-1931}
H.~Bethe, Z.\ Phys.\ {\bf 71}, 205 (1931). 

\bibitem{mossel-2012}
J.~Mossel and J.-S.~Caux, J.\ Phys.\ A:\ Math.\ Theor.\ {\bf 45}, 
255001 (2012). 


\bibitem{pozsgay-2013}
B.~Pozsgay, J.\ Stat.\ Mech.\ (2013) P07003. 

\bibitem{kor-book}
V.~E.~Korepin, N.~M.~Bogoliubov, and A.~G.~Izergin, \emph{Quantum 
Inverse Scattering Methods and Correlation Functions}, Cambridge 
University Press, Cambridge, 1997. 

\bibitem{landau-binder}
D.~P.~Landau and K.~Binder, \emph{A Guide to Monte Carlo Simulations in 
Statistical Physics}, Cambridge University Press, Cambridge, 2000.

\bibitem{faddeev-1996}
L.~D.~Faddeev, arXiv:9605187.

\bibitem{yang-1969}
C.~N.~Yang and C.~P.~Yang, J.\ Math.\ Phys. {\bf 10}, 1115 (1969).

\bibitem{creutz-1983}
M.~Creutz, Phys.\ Rev.\ Lett.\ {\bf 50}, 1411 (1983).

\bibitem{lustig-1998}
R.~Lustig, J.\ Chem.\ Phys.\ {\bf 109}, 8816 (1998).

\bibitem{prep}
In preparation. 



\end{thebibliography}
\end{document}